\input{aipcheck}

\documentclass[
    ,final            
  ]
  {aipproc}
\usepackage{amssymb}

\layoutstyle{6x9}

\begin{document}

\title{Radio Searches for Pulsars \\ and Short-Duration Transients}


\author{Maura McLaughlin}{
  address={Department of Physics, West Virginia University, Morgantown, WV 26506, USA\\ Also adjunct at the National Radio Astronomy Observatory, Green Bank, WV 24944, USA}
}

\begin{abstract}
I discuss  methods and current software packages for radio searches for pulsars and short-duration transients. I then describe the properties of the current pulsar population and the status of and predictions for ongoing and future surveys.
 The presently observed pulsar population numbers around 2000 and is expected to roughly double over the next five years, with the number of millisecond pulsars expected to more than triple. Finally, 
I  discuss individual objects discovered in the Green Bank Telescope 350-MHz Drift-Scan Survey and the Arecibo Pulsar ALFA Survey.
\end{abstract}

\maketitle




\section{Search Algorithms}

Radio searches for pulsars generally rely on two types of algorithms - periodicity searches and single-pulse searches.
The first step in either of these algorithms is `de-dispersion', or correction for the frequency-dependent delays
incurred as
radio pulses travel through the interstellar medium. Filterbank data (i.e. data recorded
as intensity vs. radio observing frequency
and time) are dedispersed at 
a number of trial dispersion measures (DMs) by appropriately shifting the frequency channels in  time and summing them to
form a `dedispersed time series'.
The range of trial DMs is typically chosen based on a model for electron density
in the Galaxy \citep{cl02}. The spacing is chosen given some minimum detectable pulse width, with the caveat that the error due to
an incorrect DM is not less than the dispersion smearing across an individual frequency channel.

 In a periodicity search, a Fast Fourier Transform (FFT) is
 applied to each dedispersed time series
to create a power or amplitude spectrum. Multiple harmonics, typically up to 16, 
are summed to increase sensitivity to narrow pulses. The time series are then folded at the periods of any resultant candidates to produce
pulse profiles. These profiles are typically inspected vs. frequency and time to gauge the reality of signals.
The signal-to-noise or $\chi^2$ values are the metrics typically used to gauge significance. 
In a single-pulse search, the dedispersed time series are analyzed by searching for any individual points in the series at which
the radio intensity is above some threshold. The time series are iteratively smoothed and searched again 
 to increase sensitivity to
broadened pulses. Events are plotted vs. DM and time and diagnostic plots are created. Example output
from each type of search is  shown in Figure~1. For further details, see \citep{lk05}. 

There are two publicly-available search packages routinely used for pulsar searching -  SIGPROC\footnote{\texttt{http://sigproc.sourceforge.net}} (with routines \texttt{dedisperse} and \texttt{seek}) and PRESTO\footnote{\texttt{https://github.com/scottransom/presto}} (with routines \texttt{prepdata} or \texttt{prepsubband} and \texttt{accelsearch}). The de-dispersion routines  \texttt{dedisperse} and \texttt{prepdata} use similar `brute-force' algorithms. For large numbers of DMs and/or large numbers of frequency channels, \texttt{prepsubband} is more efficient as it dedisperses the data in subbands (i.e. subsets of the total bandpass) and then shifts and adds those subbands to form 
 final dedispersed time series. All three routines can read in a list of frequency channels to ignore and have an option to clip the data to remove impulsive radio frequency interference (RFI). The PRESTO routine \texttt{rfifind} will output a `mask' file that can be read in with either \texttt{prepdata} or \texttt{prepsubband}. Both \texttt{seek} and \texttt{accelsearch} perform Fourier-domain periodicity searches and have algorithms for acceleration searches (necessary to correct for binary orbital motion).
 The \texttt{accelsearch} routine uses a Fourier-domain acceleration search technique \citep{rem02}, while the  \texttt{seek} routine
 uses a time-domain technique that involves `stretching' and `squeezing' the time series \citep{clf+00}.
Both \texttt{accelsearch} and \texttt{seek} can read in a file containing typical RFI frequencies to ignore in the final reported candidate list.
Both search packages also include routines for single-pulse searching. The  \texttt{seek} routine has a single-pulse search option incorporated \citep{cm03}. The PRESTO package has a separate routine named \texttt{single\_pulse\_search.py}. The algorithm used is very similar, although in \texttt{seek} adjacent pulses are added to increase sensitivity to broadened
pulses whereas in \texttt{single\_pulse\_search.py}, time series are convolved with a boxcar function.

\begin{figure}
  \includegraphics[height=.4\textheight]{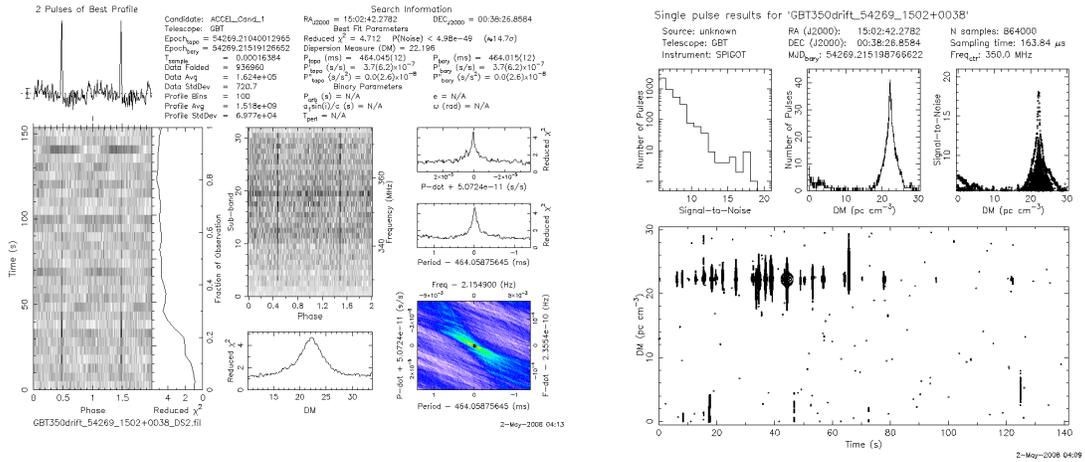}
  \caption{Diagnostic output produced by PRESTO for periodicity (left) and single-pulse (right) searches. Both plots are for the same pulsar, discovered in the GBT 350-MHz Drift-Scan Survey. The periodicity search plot shows the folded pulse profile (top left); the folded pulse profile vs. time (bottom left); the folded pulse profile vs. frequency (middle top); and the $\chi^2$ of the folded pulse profile vs. DM (middle bottom), period (right top), period derivative (right middle), and period and period derivative together (right bottom). The single-pulse search plot shows the number of pulses vs. signal-to-noise (top left); the number of pulses vs. DM (top middle); signal-to-noise vs. DM for all pulses (top right); and all pulses plotted vs. DM and time, with the size of the plotted symbol corresponding to the signal-to-noise.}
\end{figure}

\section{Current Status}

Figure~2 illustrates the history of pulsar discoveries. There are currently (as of January 2011)
 1880 pulsars listed in the ATNF pulsar
database \citep{mhth05}, with over 800 of these discovered in the Parkes Multibeam Pulsar Survey \citep{mlc+01,emk+10}.
Monte Carlo population syntheses
 suggest that there are roughly 120,000 detectable radio pulsars in the Galaxy \citep{fk06},
indicating that we have discovered less than 2\% of all possible pulsars.
The very first pulsars (i.e. the $\sim$40 detected in 1968 and 1969)
 were discovered through their bright single pulses \cite{hbp+68}.  However,
it was soon realized that most pulsars are detected with higher significance through searches for periodicity.  For that reason
most pulsar searches over the past 40 years have used solely Fourier techniques, with over 1800 of the pulsars
listed in the ATNF pulsar database  discovered through periodicity searches.
Therefore, despite some successes (e.g. \citep{nic99}), single-pulse searches were not routinely performed until
the discovery of a new class of pulsars called Rotating Radio Transients (RRATs) in 2006 \citep{mll+06}.
These objects are neutron stars that have very sporadic emission that renders them {\it only} detectable through
searches for single pulses. Giant-pulsing pulsars (e.g. \citep{cbh+04}) may also  be only detectable through single-pulse searches \citep{cm03}.
Given the small amount of extra computing time (typically < 10\% of total) necessary for single-pulse searches, they
are now routinely incorporated as a part of major pulsar searches.

\begin{figure}
  \includegraphics[height=.3\textheight]{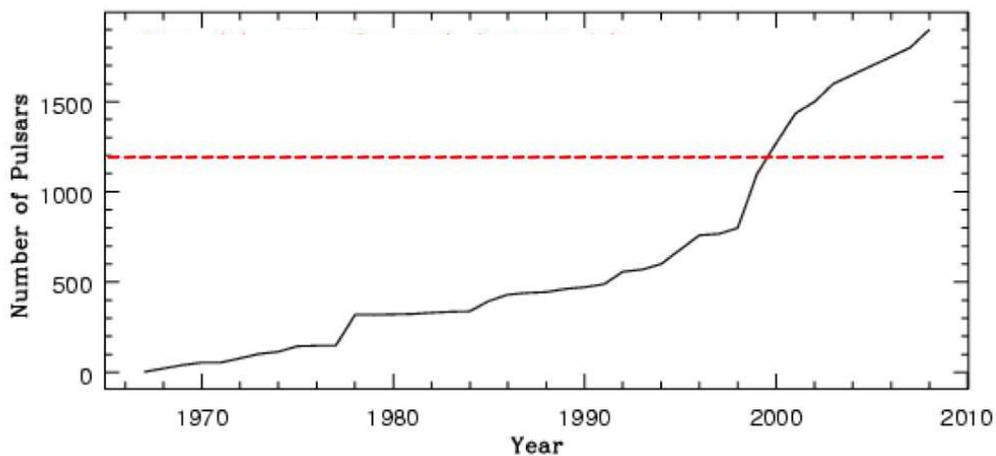}
  \caption{The number of known radio pulsars vs. year of discovery, with a dotted line indicating 1\% of the total estimated detectable
neutron star population \citep{fk06}.}
\end{figure}

Figure~3 illustrates the spin-down properties of the current pulsar population. Of the 1880 pulsars currently listed in the
ATNF pulsar database, 140 are in binary systems. Of these, $\sim$120 have white dwarf companions and $\sim$100 are fully recycled millisecond pulsars (MSPs)
with periods less than 20~ms. Half of these white dwarf binary systems are in globular clusters. All of the Galactic systems, with the exception of the unusual J1903+0327 \citep{crl+08,fbw+10}, have circular orbits while more than 10\% of the globular cluster systems are in eccentric ($e > 0.1$) orbits. Of the remaining binary pulsars, four have main sequence companions and 10 have neutron star companions. The main sequence binaries all have very eccentric ($e > 0.5$) orbits while the double neutron star eccentricities range
from 0.08 to 0.8. 

Some other classes of neutron stars are indicated on this diagram. There are over 30 known RRATs, or objects that were discovered only through their single pulses \citep{ekl09,kle+10,bb10,dcm+09,hrk+08}. Nine of those have measured period derivatives \citep{mlk+09,lmk+09,boyles11} which indicate higher magnetic fields than normal pulsars \citep{mlk+09}. 
 Magnetars, or soft gamma-ray repeaters and anomalous X-ray pulsars, are also shown. Three of these have been found to emit radio pulsations, with one of these discovered in a radio survey \citep{crh+06,crhr07,ljk+08}.
Finally, period derivatives have been measured for three X-ray detected isolated neutron stars (INSs) \citep{kk09}; no radio
pulsations have been detected from INSs, but this could easily be attributed to small beaming angles, as expected given  their long periods \citep{kml+09}.
It is unclear how the neutron star populations in the top-right part of the $P-\dot{P}$ diagram relate. One RRAT and several normal pulsars have magnetic fields comparable to magnetars. Two RRATs lie in a region of $P-\dot{P}$ space closer to INSs and some lie within the bulk of the normal  population.

Of the 1880 pulsars listed in the ATNF pulsar database, 21 are extragalactic; all of these extragalactic pulsars are located in the Small and Large Magellanic Clouds.

\begin{figure}
\includegraphics[height=.5\textheight,angle=270]{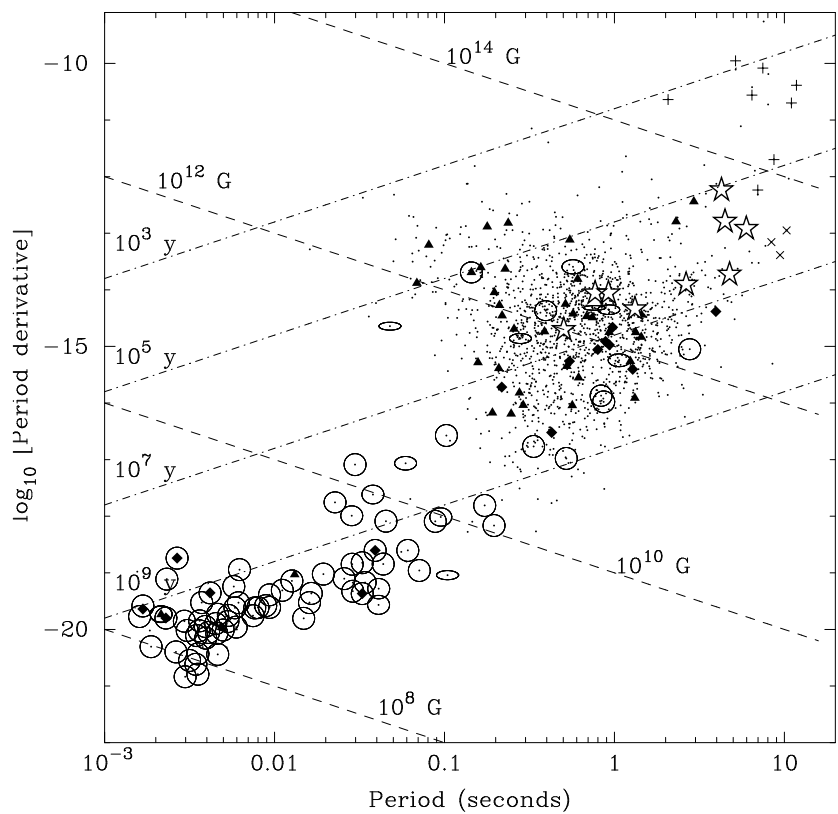}
  \caption{Period vs. period derivative for all known pulsars. Pulsars are plotted as dots, with those discovered in the PALFA/GBT 350-MHz Drift-Scan Survey plotted as diamonds/triangles. Binary pulsars are marked with circles, with the ellipticity of the circle corresponding to the ellipticity of the binary orbit. RRATs are denoted by open stars, SGRs by plus signs, and INSs by crosses. Lines of constant inferred surface dipole magnetic field strength and characteristic age are shown. }
\end{figure}

\section{Current Surveys}

Several ongoing or upcoming surveys promise to dramatically increase the known pulsar population and reveal interesting
individual objects.
We discuss the parameters of these surveys and the populations they probe here in order of observing frequency. Unless otherwise referenced, survey yields have been calculated using the PSRPOP software\footnote{\texttt{http://psrpop.phys.wvu.edu}}\citep{rl10}.

\begin{description}

\item[LOFAR 140-MHz All-Sky Survey]
LOFAR is a wide field-of-view interferometer  being commissioned in the Netherlands, with over 30 antennas currently operational, including one of the UK and one in Sweden \citep{hsa+10}. At a very low center frequency of 140~MHz, with 26~MHz of bandwidth and 64-min integrations, an all-sky survey  would discover roughly 900 pulsars \citep{ls10}, including
nearly all of the detectable pulsars within $\sim$1~kpc.
LOFAR will also be a powerful probe of pulsars in globular clusters and  local group galaxies.
\item[GBT 350-MHz Drift-Scan Survey]
Over the (Northern) summer and fall of 2007, 10,300 deg$^2$ between declinations of --21 and +26 were covered
in `drift-scan' mode during the Green Bank Telescope (GBT) track repair, with an effective integration time of 150~s.
A center frequency of 350 MHz, 50~MHz bandwidth,
2048 frequency channels, and 81.92~$\mu$s sampling time were used.  A total of 26 pulsars, including five MSPs and one RRAT, have been discovered, with  processing roughly 80\% complete\footnote{\texttt{http://www.as.wvu.edu/\textasciitilde pulsar/GBTdrift350/}}. A portion of the data (20 Terabytes, or 1500  deg$^2$)  has been
set aside for analysis by students in the Pulsar Search Collaboratory (PSC) project \citep{rhm+10}. The students have  discovered two
 pulsars and several more are expected. 
\item[GBT 350-MHz Northern Celestial Cap Survey]
The primary motivation of this survey is to find  MSPs to include in the pulsar
timing array for gravitational wave detection. This GBNCC survey has recently begun and will cover  declinations greater than 38$^\circ$ at 350 MHz with two-min integrations using  100 MHz bandwidth,
4096 frequency channels, and 81.92~$\mu$s sampling time. It
 is expected to discover  over 200 normal pulsars and several tens of MSPs.
\item[GMRT 330- and 610-MHz Surveys] Three  pulsars have been discovered in  a
 preliminary survey with the Giant Metrewave Radio Telescope in India covering $|b| < 1^\circ$ 
and $45^\circ < l < 135^\circ$ \citep{jml+09} and using a frequency of 610~MHz, with 16-MHz bandwidth,  256-$\mu$s sampling time, and 35-min integrations.
  Up to 100 pulsars (including $\sim$5 MSPs) could be discovered in a planned 350-MHz
survey with 32-MHz bandwidth $1^\circ < |b| < 10^\circ$.
The GMRT has the ability to image survey fields-of-view concurrently with the search observations, facilitating  pulsar
localization.
\item[Arecibo 1.4-GHz Pulsar ALFA Survey] This ongoing seven-beam 
 survey covers
 $|b| < 5^\circ$ and  $40^\circ < l < 75^\circ$,  $170^\circ < l < 210^\circ$ with 300 MHz of bandwidth, 256 frequency channels,  64-$\mu$s sampling time, and 265-s integrations \citep{cfl+06}. So far, 56 pulsars, including eight MSPs and six RRATs, have been found\footnote{\texttt{http://www.naic.edu/\textasciitilde palfa/newpulsars/}}. Roughly 300 pulsars, including about 50 MSPs, are expected to be
discovered.
\item[Parkes 1.4-GHz High Time Resolution Universe Survey ]
 This survey will cover the entire sky visible from Parkes with 340-MHz bandwidth, 1024 frequency channels, and 64-$\mu$s sampling time. Integrations of 4300, 540, and 270 s will be used at $|b| < 3.5^\circ$, $3.5 < |b| < 15^\circ$, and high latitudes, respectively.
 The mid-latitude part of the survey has begun and has discovered 27 pulsars, including five MSPs \citep{kjv+10}. The full survey
 is expected
to discover $\sim$400  pulsars, including 75 MSPs.
\item[Effelsberg 1.4-GHz Multibeam Survey] This survey will cover the same latitude ranges as Parkes with 1500 s, 3~min, and 1.5~min integrations with a seven-beam receiver identical to that used for  PALFA. The survey will use a bandwidth of 240 MHz, 1024 frequency channels, and 54-$\mu$s sampling time.
Roughly 600  pulsars are expected to be discovered, with $\sim$100 likely to be MSPs \citep{lfl+06}.
\item[Parkes 6-GHz Galactic Center Survey]
This recently completed survey covered $|b| < 0.25^\circ$ and  $-60^\circ < l < 30^\circ$ using a seven-beam receiver with a 576-MHz bandwidth, 192 frequency channels, and 125-$\mu$s sampling time and integration times of 1055~s. The survey discovered  
three  pulsars, all with DMs greater than 900~pc~cm$^{-3}$ \cite{bjl+10}. 
\end{description}

 A total of 4500 pulsars are predicted to be discovered by the above surveys combined.
Even accepting that half of these  pulsars will be discovered in multiple surveys, we
still expect the pulsar population to roughly double over the
next five years. This will enable precise population analyses, studies of pulsar emission phenomenology, and a clearer understanding
of how radio pulsars are related to other classes of neutron stars. Given the numbers of RRATs detected in current surveys, we expect the number of these objects to increase dramatically, to at least several hundred objects.
 This will allow us to perform population analyses to clarify how they are related to normal pulsars. The Galactic MSP population, which currently numbers around
100 (including the {\it Fermi} discoveries, e.g. \citep{rrc+11}), will more likely {\it triple} over the next five years. 
This will nearly double the sensitivity of the pulsar timing array for gravitational wave detection \citep{haa+10}.

\section{GBT 350-MHz Drift-Scan Survey}

The discoveries from this survey are marked in Figure~3 and include
J1023+0038, the first MSP showing evidence for  recent accretion  \citep{asr+09} 
and a unique probe  of  the MSP recycling process. Another highlight is the `black-widow' pulsar J2256--10,
a 2-ms eclipsing pulsar  with a minimum companion mass of only 0.03~M$_{\odot}$ \citep{stairs11}.
 We have recently detected gamma-ray pulsations with {\it Fermi}  from this source.
 A third notable discovery is
 J2222--01, a 33-ms partially recycled pulsar with a massive (i.e. minimum mass of 1.1~M$_{\odot}$) white dwarf companion \citep{boyles11}. This object seems to belong to the rare class of intermediate mass binary pulsars \citep{fsk+10}. It has a DM of only 3.27~pc~cm$^{-3}$, yielding an inferred distance \citep{cl02} of only 300~pc. This is  the second closest pulsar binary and will be an excellent target for high-precision astrometry and multi-wavelength companion characterization. A final interesting object is  J0348+04, which is only partially recycled (i.e. period of 39~ms) yet has a very low mass companion (i.e. minimum mass of 0.08~M$_{\odot}$) \citep{lynch11}. 
This object is similar to
 pulsar J1744--3922, discovered with Parkes \cite{brr+07}; together they may represent a new class of neutron star binary system.

\section{Pulsar ALFA Survey}

The discoveries from this survey are also marked in Figure~3 and include the intriguing 
 J1903+0327,
a fully recycled (2.1~ms) MSP in an eccentric ($e=0.44$) orbit with a solar mass companion \citep{crl+08}. The companion is a main-sequence star, suggesting that the pulsar was originally a member of a triple system in which the donor star was either ablated or ejected \citep{fbw+10}. The pulsar has a very high and  unusually well-determined  mass of 1.667$\pm$0.021~M$_{\odot}$. The survey has also discovered a new double neutron star system, J1906+0746 \cite{lsf+06}, in which the younger, non-recycled pulsar is the detectable source.
 This is 
the youngest binary pulsar system known.
Seven pulsars have been discovered through the single-pulse search, including one mysterious source (J1928+15) which showed three consecutive detectable pulses but  none since, despite multiple follow-up attempts
 \citep{dcm+09}.
Some of the survey data are being analyzed through the Einstein@Home distributed computing project\footnote{\texttt{http://einstein.phys.uwm.edu}}\cite{kab+10}. This has so far resulted in the discovery of two new pulsars. Both of these pulsars
are recycled and one may have a neutron star companion.

\begin{theacknowledgments}
I gratefully acknowledge my collaborators on the GBT 350-MHz Drift-Scan and PALFA survey teams, Duncan Lorimer for prediction of survey yields using the PSRPOP software, and the ASTRONS2010 organizers for the conference invitation and the travel support. This work has been supported by WVEPSCOR, the Sloan Foundation, the Research Corporation, and the NSF.

\end{theacknowledgments}



\bibliographystyle{aipproc}   

\bibliography{journals,psrrefs,modrefs}

\IfFileExists{\jobname.bbl}{}
 {\typeout{}
  \typeout{******************************************}
  \typeout{** Please run "bibtex \jobname" to optain}
  \typeout{** the bibliography and then re-run LaTeX}
  \typeout{** twice to fix the references!}
  \typeout{******************************************}
  \typeout{}
 }

\end{document}